\documentclass[twocolumn,10pt]{IEEEtran} 
\usepackage{amsmath,amsfonts}
\usepackage{algorithmic}
\usepackage{algorithm}
\usepackage{array}
\usepackage[caption=false,font=normalsize,labelfont=sf,textfont=sf]{subfig}
\usepackage{textcomp}
\usepackage{stfloats}
\usepackage{url}
\usepackage{verbatim}
\usepackage{graphicx}
\usepackage{cite}
\usepackage{setspace}
\usepackage{xcolor}
\begin{document}
\bstctlcite{BSTcontrol}

\title{Channel Estimation in MIMO Systems Aided by Microwave Linear Analog Computers (MiLACs) }

\author{Qiaosen Zhang, Matteo Nerini,~\IEEEmembership{Senior Member,~IEEE}, Bruno Clerckx,~\IEEEmembership{Fellow,~IEEE}
    
\thanks{Qiaosen Zhang, Matteo Nerini, and Bruno Clerckx are with the Department of Electrical and Electronic Engineering, Imperial College London, SW7 2AZ London, U.K (e-mail: qiaosen.zhang23@imperial.ac.uk; m.nerini20@imperial.ac.uk; b.clerckx@imperial.ac.uk).}}

% The paper headers
\markboth{}%
{}

\maketitle

%\doublespacing

\begin{abstract}
Microwave linear analog computers (MiLACs) have recently emerged as a promising solution for future gigantic multiple-input multiple-output (MIMO) systems, enabling beamforming with greatly reduced hardware and computational cost. However, channel estimation for MiLAC-aided systems remains an open problem. Conventional least squares (LS) and minimum mean square error (MMSE) estimation rely on intensive digital computation, which undermines the computational advantage offered by MiLACs. In this letter, we propose efficient LS and MMSE channel estimation schemes for MiLAC-aided MIMO systems. 
By designing the training precoder and combiner implemented by lossless and reciprocal MiLACs, the proposed schemes perform LS and MMSE estimation in the analog domain, leaving only simple digital scaling. They achieve identical estimation performance to their digital counterparts while significantly reducing computational complexity.
Numerical results verify the effectiveness of the proposed schemes.
\end{abstract}


\section{Introduction}
\label{Sec.1}
% Future Demand & Problem 
\IEEEPARstart{O}{ver} the past decades, wireless networks have evolved toward large-scale multiple-input multiple-output (MIMO). This trend is expected to continue in sixth-generation (6G) with the advent of gigantic MIMO employing thousands of antennas \cite{S1:2}. However, such scaling makes conventional digital beamforming increasingly impractical, as it requires one dedicated radio-frequency (RF) chain per antenna, incurring prohibitive hardware cost, power consumption, and computational complexity.

% MiLAC as a potential soultion 
To tackle this, recent research has focused on beamforming strategies that shift computation from the digital domain to the more efficient analog domain.
Among these strategies, microwave linear analog computer (MiLAC)-aided beamforming has recently emerged as a compelling candidate \cite{S3:4,S3:5}. 
A MiLAC is a multiport microwave network composed of tunable admittance components that processes microwave signals entirely in the analog domain. By configuring these admittance components, a MiLAC can realize a desired linear transformation between its input and output signals.
As a result, MiLAC-aided beamforming offers flexibility comparable to digital beamforming while requiring only a minimal number of RF chains and low-resolution analog-to-digital/digital-to-analog converters (ADCs/DACs).
In particular, it substantially reduces computational complexity by realizing the desired linear transformation through wave propagation in the MiLAC, rather than through arithmetic operations in the digital domain.

% Literature Review + Gap + Limitation & Challenge in CE
Existing studies on MiLAC have focused on modelling \cite{S3:4}, beamforming design \cite{S2:3,S3:5,MiLAC_zheyu}, and architecture design \cite{S2:4}, while channel state information (CSI) acquisition remains an open problem. Accurate CSI is essential for MiLAC-aided beamforming, as it determines the configuration of tunable admittance components.
Although MiLACs can efficiently perform pre-multiplication of input signals by precoders or combiners in the analog domain, thereby enabling MiLAC-aided beamforming, this is insufficient for conventional least squares (LS) and minimum mean square error (MMSE) channel estimation. This is because conventional LS and MMSE require aggregating received signals over multiple training time slots and processing the resulting matrix via operations such as post-multiplication and vectorization \cite{S3:1}, which necessitate digital storage and computation beyond the capabilities of MiLACs. Consequently, achieving channel estimation with MiLACs without compromising the computational advantage of MiLAC-aided beamforming is a non-trivial problem.

% Contributions
To address this gap, in this letter, we investigate channel estimation in a point-to-point MIMO system aided by 
lossless and reciprocal MiLACs. The main contributions are as follows.
\textit{First}, we model a MiLAC-aided MIMO system, where the transmitter- and receiver-side MiLACs implement the training precoder and combiner, respectively, for channel estimation.
\textit{Second}, we propose efficient LS and MMSE channel estimation schemes for the MiLAC-aided system. 
By jointly designing the training precoder and combiner along with the source signals, the columns of the channel estimate can be obtained from the receive RF chains up to digital scaling. This enables LS and MMSE estimation to be carried out in the analog domain, achieving identical performance to their digital counterparts while significantly reducing computational complexity.
\textit{Third}, we present numerical results on normalized mean square error (NMSE) and computational complexity, which verify the effectiveness of the proposed schemes.

\section{System Model}
\label{Sec.2}
\begin{figure}[!t]
\centering
\includegraphics[width=1\linewidth]{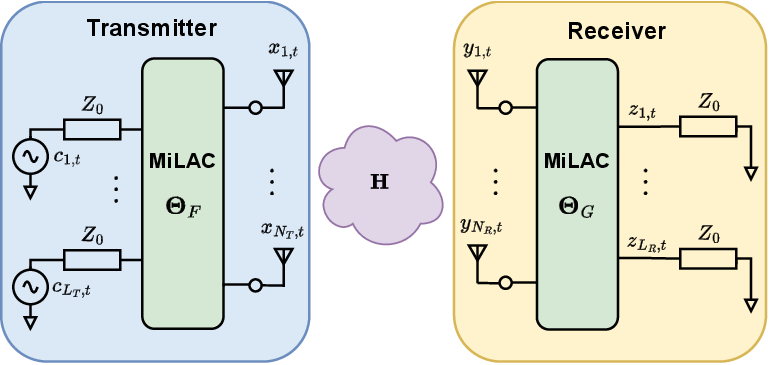}
\caption{MiLAC-aided MIMO system model.}
\label{fig.1}
\end{figure}

% General Set up 
We consider a narrowband point-to-point MIMO system aided by a MiLAC at both the transmitter and receiver, as shown in Fig.~\ref{fig.1}. The $N_T$-antenna transmitter and $N_R$-antenna receiver are equipped with $L_T$ and $L_R$ RF chains, respectively, where $L_T \leq N_T$ and $L_R \leq N_R$. 
Following \cite{S2:1}, each transmit RF chain and receive antenna is modelled as a voltage source with a source impedance $Z_0$ (typically $50~\Omega$), while each transmit antenna and receive RF chain is modelled as a terminal loaded with $Z_0$ 
\footnote{For analytical tractability, we consider power matching in this letter and leave noise matching, which is more common in receiver RF design, for future work.}.
We assume no mutual coupling among antennas at either end and adopt the unilateral approximation between the transmitter and receiver \cite{S2:1}.

% Transmitter 
As shown in Fig.~\ref{fig.1}, at the $t$-th training time slot, $\forall t \in \{1, \dots, \tau\}$, the source signal at transmit RF chains is denoted as $\mathbf{c}_t = \bigl[c_{1,t},\dots,c_{L_T,t}\bigr]^T \in \mathbb{C}^{L_T \times 1}$ 
\footnote{In this letter, the signals at the MiLAC input and output are RF signals represented by the corresponding discrete-time samples of their complex baseband equivalents. They are interpreted as root-mean-square (RMS) voltages.}. The source signal $\mathbf{c}_t$ is precoded by the transmitter-side MiLAC, generating the training signal $\mathbf{x}_t = \bigl[x_{1,t},\dots,x_{N_T,t}\bigr]^T \in \mathbb{C}^{N_T \times 1}$ at the \(N_T\) antennas, 
written as $\mathbf{x}_t=\mathbf{F}\mathbf{c}_t$, where $\mathbf{F} \in \mathbb{C}^{N_T\times L_T}$
denotes the training precoder implemented by the transmitter-side MiLAC. 
Referring to \cite{S2:4}, the precoder $\mathbf{F}$ 
can be expressed in terms of the scattering matrix of the transmitter-side MiLAC $\mathbf{\Theta}_{F} \in \mathbb{C}^{(L_T+N_T)\times(L_T+N_T)}$ as 
\begin{equation}
\label{eq.SM.1}
    \mathbf{F} = \frac{1}{2}[\mathbf{\Theta}_{F}]_{L_T+\mathcal{N}_T, \mathcal{L}_{T}} \;,
\end{equation}
where $[\cdot]_{L_T+\mathcal{N}_T, \mathcal{L}_{T}}$ extracts the submatrix composed of the rows and columns with indices $L_T+\mathcal{N}_T = \{L_T+1, \dots, L_T+N_T\}$ and $\mathcal{L}_{T} = \{1, \dots, L_T\}$, respectively. 

% Receiver 
The signal received at the $N_R$ antennas is denoted as $\mathbf{y}_t =\bigl[y_{1,t},\dots,y_{N_R,t}\bigr]^T \in \mathbb{C}^{N_R \times 1}$, written as $\mathbf{y}_t = \mathbf{H}\mathbf{x}_t+\mathbf{n}_t$, where $\mathbf{H} \in \mathbb{C}^{N_R \times N_T}$ is the MIMO channel between transmitter and receiver, which is assumed to be quasi-static, i.e., the channel $\mathbf{H}$ remains constant within a coherence block \cite{S2:5}. The term $\mathbf{n}_t \in \mathbb{C}^{N_R \times 1}$ denotes the additive white Gaussian noise (AWGN), with $\mathbf{n}_{t}\sim\mathcal{CN}(\mathbf{0},\sigma^2\mathbf{I}_{N_R})$, where $\sigma^2$ is the noise variance. The signal $\mathbf{y}_t$ is processed by the receiver-side MiLAC, giving the signal at the receive RF chains $\mathbf{z}_t = \bigl[z_{1,t},\dots,z_{L_R,t}\bigr]^T \in \mathbb{C}^{L_R \times 1}$, written as $\mathbf{z}_t=\mathbf{G}\mathbf{y}_t$, where $\mathbf{G} \in \mathbb{C}^{L_R \times N_R}$
denotes the training combiner implemented by the receiver-side MiLAC.
According to \cite{S2:4}, the combiner $\mathbf{G}$
can be expressed in terms of the scattering matrix of the receiver-side MiLAC $\mathbf{\Theta}_{G} \in \mathbb{C}^{(N_R+L_R)\times(N_R+L_R)}$ as
\begin{equation}
\label{eq.SM.2}
    \mathbf{G} = \frac{1}{2}[\mathbf{\Theta}_{G}]_{N_R+\mathcal{L}_{R},\mathcal{N}_R} \;,
\end{equation}
where $[\cdot]_{N_R+\mathcal{L}_{R},\mathcal{N}_R}$ extracts the submatrix composed of the rows and columns with indices $N_{R}+\mathcal{L}_{R} = \{N_{R}+1, \dots, N_{R}+L_R\}$ and $\mathcal{N}_{R} = \{1, \dots, N_{R}\}$, respectively. 
Finally, the end-to-end system model at the $t$-th time slot is given by  
\begin{equation}
\label{eq.SM.3}
\mathbf{z}_t = \mathbf{G}\mathbf{H}\mathbf{F}\mathbf{c}_t+\mathbf{G}\mathbf{n}_t.
\end{equation}

% Stress the constraint on MiLAC 
In this letter, we consider lossless and reciprocal MiLACs \cite{S2:4}, whose scattering matrices are assumed to be fully reconfigurable subject to
\begin{align}
    \label{eq.SM.constraint_F}
    \mathbf{\Theta}_F^H\mathbf{\Theta}_F &= \mathbf{I}_{L_T+N_T}, \mathbf{\Theta}_F=\mathbf{\Theta}_F^T, \\
    \label{eq.SM.constraint_G}
    \mathbf{\Theta}_G^H\mathbf{\Theta}_G &= \mathbf{I}_{N_R+L_R}, \mathbf{\Theta}_G=\mathbf{\Theta}_G^T.
\end{align}
Under these constraints, we design the training precoder $\mathbf{F}$ and combiner $\mathbf{G}$ along with the source signal $\mathbf{c}_t$ for channel estimation. 
We assume $L_T=N_T$ and $L_R=N_R$, and design $[\mathbf{\Theta}_{F}]_{L_T+\mathcal{N}_T, \mathcal{L}_{T}} = 2\mathbf{F}$ and $[\mathbf{\Theta}_{G}]_{N_R+\mathcal{L}_{R},\mathcal{N}_R} = 2\mathbf{G}$ to be unitary. With this design, the transmit power, i.e., the power dissipated by the transmit antennas, defined as $P_T= \sum_{t=1}^{\tau}\|\mathbf{x}_t\|^2/Z_0$, is equal to the available power of the sources, given by $\sum_{t=1}^{\tau}\|\mathbf{c}_t\|^2/(4Z_0)$ \cite[Section~12.1]{Microwave_Engineering_Book}.
Finally, to ensure the $N_RN_T$ unknown entries of $\mathbf{H}$ are resolvable with minimal training overhead, we set $\tau=N_T$.

\section{MiLAC-aided LS Channel Estimation}
\label{Sec.LS}
In this section, we address LS channel estimation for the MiLAC-aided MIMO system described in Section \ref{Sec.2}. 
Our goal is to perform LS estimation in the analog domain, with only a digital constant scaling.
This is non-trivial since conventional LS estimation requires post-multiplying the received signal matrix by the LS estimator, which is beyond the computational capabilities of MiLACs. To address this, we jointly design the source signal $\mathbf{c}_t$, training precoder $\mathbf{F}$, and training combiner $\mathbf{G}$, such that the $t$-th column of the LS estimate is directly obtained from $\mathbf{z}_t$ up to a constant scaling. We then show that the resulting $\mathbf{F}$ and $\mathbf{G}$ are physically implementable by MiLACs under \eqref{eq.SM.constraint_F} and \eqref{eq.SM.constraint_G}.

\subsection{Training Precoder and Combiner Design}
% Form the received signal matrix & Show the general LS estimator 
We begin by recalling that the received signal at the $t$-th time slot is expressed as $\mathbf{y}_t=\mathbf{H}\mathbf{x}_t+\mathbf{n}_t$. By collecting $\mathbf{y}_t$ over the $\tau$ time slots, we obtain the received signal matrix $\mathbf{Y}=[\mathbf{y}_1,\dots,\mathbf{y}_\tau] \in \mathbb{C}^{N_R\times\tau}$ as  
\begin{equation}
\label{eq.LS.1}
    \mathbf{Y} = \mathbf{H}\mathbf{X}+\mathbf{N},
\end{equation}
where $\mathbf{X} = [\mathbf{x}_1,\dots,\mathbf{x}_\tau] \in \mathbb{C}^{N_T\times\tau}$ is the training matrix satisfying $\|\mathbf{X}\|_F^2 = \bar{P}_T$, with $\bar{P}_T = P_TZ_0$, and $\mathbf{N} = [\mathbf{n}_1,\dots,\mathbf{n}_\tau] \in \mathbb{C}^{N_R\times\tau}$ is the noise matrix. The LS estimate of $\mathbf{H}$ is given by
\begin{equation}
\label{eq.LS.2}
    \hat{\mathbf{H}}_{\text{LS}} = \mathbf{Y}\mathbf{X}^H(\mathbf{X}\mathbf{X}^H)^{-1}.
\end{equation}

% Show the optimal training matrix & our choice to simplify the estimator 
To minimize the mean square error (MSE) of (\ref{eq.LS.2}) subject to $\|\mathbf{X}\|_F^2 = \bar{P}_T$, the optimal training matrix must satisfy $\mathbf{X}\mathbf{X}^H=\bar{P}_T/N_T\mathbf{I}_{N_T}$ \cite[Section~III]{S3:1}. Without loss of generality, we choose $\mathbf{X}=\sqrt{\bar{P}_T/N_T}\mathbf{I}_{N_T}$ 
\footnote{Such a design is typically avoided in digital systems because it forces each transmit RF chain to switch between full and zero power, inducing prohibitive peak-to-average power ratio (PAPR). In contrast, this poses no issue in MiLAC-aided systems, since $\mathbf{c}_t$ can be chosen to be entry-wise constant-modulus, enabling the transmitter-side MiLAC to generate the desired $\mathbf{x}_t$ while maintaining unit PAPR at each RF chain, i.e., $\max_{t}|c_{i,t}|^2 = (1/\tau)\sum_{t=1}^{\tau} |c_{i,t}|^2, \forall i \in \{1, \dots, L_T\}$.}, 
so that (\ref{eq.LS.2}) reduces to 
\begin{equation}
\label{eq.LS.3}
    \hat{\mathbf{H}}_{\text{LS}} = \sqrt{\frac{N_T}{\bar{P}_T}}\mathbf{Y}.
\end{equation}
% Based on the choosen training matrix and simplified estimator, show how c_t can be chosen to ensure unit PAPR & how F_t can be set to achieve optimal training matrix & how G_t can be set to ensure final channel estimate
Under this choice, the training signal at the $t$-th time slot becomes $\mathbf{x}_t=\sqrt{\bar{P}_T/N_T}[\mathbf{I}_{N_T}]_{:,t}$. 
To realize such $\mathbf{x}_t$ while ensuring unit PAPR at each RF chain, we choose $\mathbf{c}_t = \sqrt{4\bar{P}_T/N_T}[\mathbf{D}]_{:,t}$, where $\mathbf{D} \in \mathbb{C}^{N_T\times \tau}$ is the normalized discrete Fourier transform (DFT) matrix satisfying $\mathbf{D}^H\mathbf{D} =\mathbf{D}\mathbf{D}^H= \mathbf{I}_{\tau}$. Accordingly, we set the training precoder 
\begin{equation}
\label{eq.LS.4}
    \mathbf{F}=\frac{1}{2}\mathbf{D}^H. 
\end{equation}
At the receiver, we simply choose the training combiner 
\begin{equation}
\label{eq.LS.5}
    \mathbf{G}=\frac{1}{2}\mathbf{I}_{N_R},
\end{equation}
such that the $t$-th column of the LS estimate is obtained from $\mathbf{z}_t$ via a constant scaling, i.e., $[\hat{\mathbf{H}}_{\text{LS}}]_{:,t} = \sqrt{4N_T/\bar{P}_T}\mathbf{z}_t$.
Notably, \eqref{eq.LS.5} reduces the receiver-side MiLAC to direct connections between the receive antennas and their corresponding RF chains. Hence, the receiver-side MiLAC can be omitted for LS estimation, whereas it plays a non-trivial role in MMSE estimation as presented in Section~\ref{Sec.MMSE}.

% Summary
In summary, the proposed design of $\mathbf{c}_t$, $\mathbf{F}$, and $\mathbf{G}$ shifts LS channel estimation into the analog domain, leaving only a digital constant scaling.

\subsection{Implementation of the Training Precoder and Combiner}
\label{sec.LS.B}
% Introduction and review main assumption 
We next show that the proposed training precoder $\mathbf{F}$ and combiner $\mathbf{G}$ are physically realizable by lossless and reciprocal MiLACs. Recall that $L_R = N_R$ and $L_T = N_T$.

% Show the corresponding scattering matrix subject to the constraints but realize F and G
To this end, we first determine scattering matrices $\mathbf{\Theta}_F$ and $\mathbf{\Theta}_G$ such that $[\mathbf{\Theta}_{F}]_{L_T+\mathcal{N}_T, \mathcal{L}_{T}} = 2\mathbf{F}$ and $[\mathbf{\Theta}_{G}]_{N_R+\mathcal{L}_{R},\mathcal{N}_R} = 2\mathbf{G}$, subject to \eqref{eq.SM.constraint_F} and \eqref{eq.SM.constraint_G}, respectively. 
Since $2\mathbf{F}$ and $2\mathbf{G}$ are unitary, the scattering matrices must take the forms
\begin{equation}
    \label{eq.LS.Theta}
        \mathbf{\Theta}_{F} = 
        \begin{bmatrix} 
            \mathbf{0}_{L_T} & 2\mathbf{F}^T\\
            2\mathbf{F} & \mathbf{0}_{N_T}\  
        \end{bmatrix},\
        \mathbf{\Theta}_{G} = 
        \begin{bmatrix} 
            \mathbf{0}_{N_R} & 2\mathbf{G}^T\\ 
            2\mathbf{G} & \mathbf{0}_{L_R}
        \end{bmatrix}.
\end{equation}
% We show the corresponding admittance matrices & mention the resulting tunbale admittance componenets
From \cite[(23)]{S2:3}, the corresponding admittance matrices $\mathbf{Y}_F \in \mathbb{C}^{(L_T+N_T)\times(L_T+N_T)}$ and $\mathbf{Y}_G \in \mathbb{C}^{(N_R+L_R)\times(N_R+L_R)}$ are related to the scattering matrices as
\begin{align}
    \label{eq.LS.YF_intermsof_ThetaF}
    \mathbf{Y}_F & = Y_0\Bigl(2\bigl(\mathbf{\Theta}_F+\mathbf{I}_{L_T+N_T}\bigr)^{-1}-\mathbf{I}_{L_T+N_T}\Bigr),\\
    \label{eq.LS.YG_intermsof_ThetaG}
    \mathbf{Y}_G & = Y_0\Bigl(2\bigl(\mathbf{\Theta}_G+\mathbf{I}_{N_R+L_R}\bigr)^{-1}-\mathbf{I}_{N_R+L_R}\Bigr),
\end{align}
where $Y_0 = 1/Z_0$ is the reference admittance. To avoid possible singularity of $(\mathbf{\Theta}_F+\mathbf{I}_{L_T+N_T})$ and $(\mathbf{\Theta}_G+\mathbf{I}_{N_R+L_R})$, we introduce $\bar{\mathbf{\Theta}}_F = e^{j\theta}\mathbf{\Theta}_F$ and $\bar{\mathbf{\Theta}}_G = e^{-j\theta}\mathbf{\Theta}_G$, where $\theta$ can be adjusted such that the corresponding inverses exist, e.g., $\theta=\pi/4$ for the current design. These phase shifts are introduced only to ensure well-defined admittance matrices for the proposed $\mathbf{F}$ and $\mathbf{G}$, which does not alter the resulting estimation performance. By substituting $\bar{\mathbf{\Theta}}_F$ and $\bar{\mathbf{\Theta}}_G$ into \eqref{eq.LS.YF_intermsof_ThetaF} and \eqref{eq.LS.YG_intermsof_ThetaG}, and applying \cite[Proposition~1]{S3:4}, we have
\begin{align}
    \label{eq.LS.YF}
    \mathbf{Y}_{F} & = Y_0
    \begin{bmatrix} 
        -2\mathbf{J}_F-\mathbf{I}_{L_T}  &  4e^{j\theta}\mathbf{J}_F\mathbf{F}^T \\
        4e^{j\theta}\mathbf{F}\mathbf{J}_F & \mathbf{I}_{N_T}-8e^{j2\theta}\mathbf{F}\mathbf{J}_F\mathbf{F}^T
    \end{bmatrix},\\
    \label{eq.LS.YG}
    \mathbf{Y}_{G} &= Y_0
    \begin{bmatrix} 
        -2\mathbf{J}_G-\mathbf{I}_{N_R}  &  4e^{-j\theta}\mathbf{J}_G\mathbf{G}^T \\
        4e^{-j\theta}\mathbf{G}\mathbf{J}_G & \mathbf{I}_{L_R}-8e^{-j2\theta}\mathbf{G}\mathbf{J}_G\mathbf{G}^T
    \end{bmatrix},
\end{align}
where $\mathbf{J}_F = (4e^{j2\theta}\mathbf{F}^T\mathbf{F}-\mathbf{I}_{L_T})^{-1}$ and $\mathbf{J}_G = (4e^{-j2\theta}\mathbf{G}^T\mathbf{G}-\mathbf{I}_{N_R})^{-1}$.
Given \eqref{eq.LS.YF} and \eqref{eq.LS.YG}, the tunable admittance components can be obtained via \cite[(7)]{S3:4}, and the corresponding circuits can then be constructed following \cite[Fig.~2]{S3:4}, thereby physically realizing the transmitter- and receiver-side MiLACs.

These results confirm that the proposed training precoder $\mathbf{F}$ and combiner $\mathbf{G}$ can be realized by lossless and reciprocal MiLACs, which proves the physical feasibility of the proposed MiLAC-aided LS channel estimation.

\subsection{Comparison with Digital LS Channel Estimation}
\label{Sec.LS.C}
% For clear comparison, describe how digital LS channel estiamtion works
To facilitate the comparison, we first describe how conventional digital LS channel estimation operates. In digital LS, the identity training matrix $\mathbf{X}=\sqrt{\bar{P}_T/N_T}\mathbf{I}_{N_T}$ used in MiLAC-aided LS cannot be adopted, because it leads to prohibitive PAPR at each transmit RF chain. For this reason, digital LS conventionally adopts a scaled DFT training matrix satisfying $\mathbf{X}\mathbf{X}^H=\bar{P}_T/N_T\mathbf{I}_{N_T}$ \cite{S3:1}. Under this choice, the LS estimator $\mathbf{X}^H(\mathbf{X}\mathbf{X}^H)^{-1}$ reduces to $N_T/\bar{P}_T\mathbf{X}^H$, rather than the identity weighting $\sqrt{N_T/\bar{P}_T}\mathbf{I}_{N_R}$ applied in MiLAC-aided LS. Consequently, digital LS must apply the LS estimator to the received signal matrix $\mathbf{Y}$ to recover the channel.

% Advantages of our MiLAC-aided LS channel estimation
Compared with digital LS, the proposed MiLAC-aided LS achieves identical estimation performance, since both use training matrices that satisfy the LS optimality condition. Moreover, it greatly reduces computational complexity. Specifically, to realize the training precoder $\mathbf{F}$ and combiner $\mathbf{G}$ as in \eqref{eq.LS.4} and \eqref{eq.LS.5}, the admittance matrices in \eqref{eq.LS.YF} and \eqref{eq.LS.YG}, together with the corresponding admittance components of MiLACs, are computed only once rather than in each coherence block. They can therefore be precomputed and are excluded from the computational complexity, as defined in the Appendix. Hence, MiLAC-aided LS involves only a constant scaling of $\mathbf{z}_t$ over the $\tau$ time slots, requiring $2\tau N_R$ real operations. In contrast, given the fixed DFT training matrix and the corresponding LS estimator, digital LS is dominated by the matrix-matrix product $\mathbf{Y}\mathbf{X}^{H}$, which requires $8\tau N_RN_T$ real operations.

\section{MiLAC-aided MMSE Channel Estimation}
\label{Sec.MMSE}
In this section, we introduce MiLAC-aided MMSE channel estimation, which improves performance over MiLAC-aided LS by exploiting channel correlation. This is challenging because conventional MMSE requires collecting received signals over the $\tau$ time slots and applying the MMSE estimator to the vectorized received signal matrix, which cannot be supported by MiLACs. To tackle this, we begin by characterizing a correlated MIMO channel with the canonical statistical model and assume that the transmit and receive correlation matrices are known. This enables MMSE estimation on the virtual channel, whose correlation matrix is diagonal. By leveraging this diagonal structure,
we jointly design the source signal $\mathbf{c}_t$, training precoder $\mathbf{F}$, and training combiner $\mathbf{G}$, so that MMSE channel estimation is performed in the analog domain, leaving only entry-wise digital scaling.

\subsection{Channel Model}
\label{Sec.MMSE.A}
% Introduce the channel model
We consider a correlated MIMO channel described by the canonical statistical model \cite{S3:2}
\begin{equation}
\label{eq.MMSE.1}
    \mathbf{H} = \mathbf{U}_R \mathbf{H}_v \mathbf{U}_T^H,
\end{equation}
where $\mathbf{U}_T \in \mathbb{C}^{N_T \times N_T}$ and $\mathbf{U}_R \in \mathbb{C}^{N_R \times N_R}$ are unitary matrices containing the eigenvectors of the transmit and receive correlation matrices, $\mathbf{R}_T = \mathbf{U}_T \mathbf{\Lambda}_T \mathbf{U}_T^H$ and $\mathbf{R}_R = \mathbf{U}_R \mathbf{\Lambda}_R \mathbf{U}_R^H$, respectively, while the diagonal matrices $\mathbf{\Lambda}_T \in \mathbb{C}^{N_T \times N_T}$ and $\mathbf{\Lambda}_R \in \mathbb{C}^{N_R \times N_R}$ capture the corresponding eigenvalues. The matrix $\mathbf{H}_v \in \mathbb{C}^{N_R \times N_T}$ denotes the virtual (or eigen-domain) channel, whose entries are zero-mean and uncorrelated with a correlation matrix given by 
\begin{equation}
\label{eq.MMSE.2}
    \mathbf{R}_v = \mathbb{E}[\mathbf{h}_v \mathbf{h}_v^H] = \mathbf{\Lambda}_T \otimes \mathbf{\Lambda}_R,
\end{equation}
where $\mathbf{h}_v = \text{vec}(\mathbf{H}_v) \in \mathbb{C}^{N_RN_T\times 1}$, and we assume $\mathbf{R}_v$ to be full rank, as commonly adopted in the literature \cite{S3:2,S3:3,S3:6}. 

% Explain why we can and want do channel estimation on Hv instead of H 
From (\ref{eq.MMSE.1}), $\mathbf{H}$ and $\mathbf{H}_v$ are unitarily equivalent, so MMSE channel estimation can be performed on $\mathbf{H}_v$ rather than $\mathbf{H}$, without any loss in the MSE \cite{S3:2}. Importantly, the diagonal structure of $\mathbf{R}_v$ greatly simplifies MMSE channel estimation, enabling the source signal $\mathbf{c}_t$, training precoder $\mathbf{F}$, and training combiner $\mathbf{G}$ to be designed so that the $t$-th column of the MMSE estimate is obtained from the signal $\mathbf{z}_t$ via entry-wise scaling, as detailed in Section~\ref{Sec.MMSE.B}.

\subsection{Training Precoder and Combiner Design}
\label{Sec.MMSE.B}
% Form the stacked received signal in virtual domain & Show the MMSE estimator 
Recall that $L_R = N_R$ and $\tau=N_T$ are assumed. Starting from the channel model in (\ref{eq.MMSE.1}), (\ref{eq.LS.1}) can be rewritten as 
\begin{equation}
\label{eq.MMSE.3}
     \mathbf{Y}_v = \mathbf{H}_v \mathbf{X}_v+\mathbf{N}_v,
\end{equation}
where $\mathbf{Y}_v = \mathbf{U}_R^H\mathbf{Y} \in \mathbb{C}^{N_R\times\tau}$, $\mathbf{X}_v = \mathbf{U}_T^H \mathbf{X} \in \mathbb{C}^{N_T\times\tau}$, and $\mathbf{N}_v = \mathbf{U}_R^H\mathbf{N} \in \mathbb{C}^{N_R\times\tau}$. By vectorizing $\mathbf{Y}_v$, we obtain  
\begin{equation}
\label{eq.MMSE.4}
     \mathbf{y}_v = \mathbf{W}\mathbf{h}_v +\mathbf{n}_v,
\end{equation}
where $\mathbf{y}_v = \text{vec}(\mathbf{Y}_v) \in \mathbb{C}^{\tau N_R \times 1}$ and $\mathbf{n}_v = \text{vec}(\mathbf{N}_v) \in \mathbb{C}^{\tau N_R \times 1}$, with $\mathbf{W} = (\mathbf{X}_v^T \otimes \mathbf{I}_{N_R}) \in \mathbb{C}^{\tau N_R \times N_RN_T}$ following from \cite[Theorem 13.26]{S3:11}. 
The MMSE estimate of $\mathbf{h}_v$ is
\begin{equation}
\label{eq.MMSE.5}
     \hat{\mathbf{h}}_{v,\text{MMSE}} = \mathbf{A}\mathbf{y}_v,
\end{equation}
where $\mathbf{A} \in \mathbb{C}^{N_RN_T \times \tau N_R}$ is given by  
\begin{equation}
\label{eq.MMSE.6}
    \mathbf{A} 
    = \mathbf{R}_{v}\mathbf{W}^H\Bigl(\mathbf{W}\mathbf{R}_{v}\mathbf{W}^H+ \sigma^2\mathbf{I}_{\tau N_R}\Bigr)^{-1}.
\end{equation}
The resulting MSE is expressed as 
\begin{equation}
\label{eq.MMSE.N1}
    \text{MSE} = \operatorname{tr}\Bigl(\bigl(\mathbf{R}_v^{-1}+\frac{1}{\sigma^2}\mathbf{W}^H\mathbf{W}\bigr)^{-1}\Bigr).
\end{equation}

% Show optimal virtual training matrix -> training signal -> Precoder Design
Given the diagonal structure of $\mathbf{R}_v$, the optimal $\mathbf{X}_v$ minimizing (\ref{eq.MMSE.N1}) subject to $\|\mathbf{X}_v\|_F^2 = \bar{P}_T$ must be diagonal \cite{S3:2}, i.e., $\mathbf{X}_v = \text{diag}(\sqrt{p_1},\dots,\sqrt{p_\tau})$, where $p_t$ denotes the power allocated to the $t$-th transmit eigen-direction. The optimal power allocation can be obtained by solving
\begin{align}
    \min_{p_t,\forall t} \quad 
    & \sum_{t=1}^{\tau} \sum_{j=1}^{N_R}
    \frac{\sigma^2[\mathbf{R}_v]_{k,k}}
    {\sigma^2 + p_t[\mathbf{R}_v]_{k,k}} \label{eq.MMSE.7} \\
    \text{s.t.}\quad 
    & \sum_{t=1}^{\tau} p_t \leq \bar{P}_T, \label{eq.MMSE.8}
\end{align}
where (\ref{eq.MMSE.7}) follows from expanding (\ref{eq.MMSE.N1}), with $k = (t-1)N_R+j$
\footnote{Since (\ref{eq.MMSE.7}) involves double summations, a closed-form solution cannot be obtained under conventional water-filling. Instead, it can be efficiently solved via a two-layer water-filling-type method derived from the Lagrange multipliers and Karush–Kuhn–Tucker (KKT) conditions, as detailed in \cite{S3:3}.}.
Since $\mathbf{X}=\mathbf{U}_T\mathbf{X}_v$, the training signal at $t$-th time slot is $\mathbf{x}_t=\sqrt{p_t}\mathbf{u}_t$, where $\mathbf{u}_t \in \mathbb{C}^{N_T\times1}$ denotes the $t$-th column of $\mathbf{U}_T$. To generate such $\mathbf{x}_t$,
we set the source signal $\mathbf{c}_t = \sqrt{4p_t}[\mathbf{D}]_{:,t}$ and the training precoder
\begin{equation}
\label{eq.MMSE.10}
    \mathbf{F}=\frac{1}{2}\mathbf{U}_T\mathbf{D}^H.
\end{equation}

% Show the simplified estimator -> Combiner design to ensure z_t = column of channel estimate
Moreover, with both $\mathbf{R}_v$ and $\mathbf{X}_v$ diagonal, the MMSE estimator in (\ref{eq.MMSE.6}) becomes diagonal, whose $k$-th diagonal entry is expressed as 
\begin{equation}
\label{eq.MMSE.9}
    [\mathbf{A}]_{k,k} 
    = \frac{\sqrt{p_t}\,[\mathbf{R}_v]_{k,k}}
    {\sigma^2 + p_t[\mathbf{R}_v]_{k,k}}.
\end{equation}
Accordingly, we choose the training combiner 
\begin{equation}
\label{eq.MMSE.11}
    \mathbf{G}=\frac{1}{2}\mathbf{U}_R^H,
\end{equation}
such that the $t$-th column of the MMSE estimate can be obtained from $\mathbf{z}_t$ via entry-wise scaling, i.e., $[\hat{\mathbf{H}}_{v,\text{MMSE}}]_{:,t} = 2\mathbf{A}_t\mathbf{z}_t$, where $\mathbf{A}_t = [\mathbf{A}]_{\mathcal{I}_t,\mathcal{I}_t} \in \mathbb{C}^{N_R \times N_R}$ with $\mathcal{I}_t =\{(t-1)N_R+1,\dots,tN_R\}$.

% Summary
Overall, the proposed design of $\mathbf{c}_t$, $\mathbf{F}$, and $\mathbf{G}$ enables MMSE channel estimation to be carried out in the analog domain, leaving only entry-wise digital scaling.

\subsection{Comparison with Digital MMSE Channel Estimation}
\label{Sec.MMSE.C}

% MiLAC admittance matrix design & Advantages 
We can readily verify that the proposed training precoder $\mathbf{F}$ and combiner $\mathbf{G}$ can be physically realized by lossless and reciprocal MiLACs. Specifically, their corresponding scattering matrices follow directly from \eqref{eq.LS.Theta}, while the resulting admittance matrices are given by \eqref{eq.LS.YF} and \eqref{eq.LS.YG}. Thus, the proposed MiLAC-aided MMSE channel estimation can achieve the same minimum MSE as its digital counterpart. Moreover, it can substantially reduce computational complexity, following reasoning similar to that in Section~\ref{Sec.LS.C}. Since the transmit and receive correlation matrices $\mathbf{R}_T$ and $\mathbf{R}_R$ vary slowly across coherence blocks \cite{S3:2,S3:3}, we assume that they are known. The MMSE estimator $\mathbf{A}$ in \eqref{eq.MMSE.6} and the admittance components of MiLACs realizing $\mathbf{F}$ and $\mathbf{G}$ can therefore be precomputed. Hence, the only digital computation of MiLAC-aided MMSE at each coherence block is the entry-wise scaling of $\mathbf{z}_t$ over the $\tau$ time slots, requiring $2\tau N_R$ real operations.
In contrast, although $\mathbf{A}$, $\mathbf{F}$, and $\mathbf{G}$ can be precomputed, digital MMSE still requires approximately $8\tau N_R^2$ real operations to compute the matrix-vector product $\mathbf{G}\mathbf{y}_t$ over the $\tau$ time slots.

\section{Numerical Results}
% Set-up
In this section, we present numerical results to evaluate the performance of the proposed MiLAC-aided LS and MMSE channel estimation and to quantify their advantages in terms of computational complexity, compared with their digital counterparts. To this end, we assume that the transmit and receive correlation matrices, $\mathbf{R}_T$ and $\mathbf{R}_R$, follow the widely adopted exponential correlation model \cite{S3:1,S3:2,S3:3}, expressed as 
$[\mathbf{R}_T]_{i,j} = \epsilon_T^{|i-j|}, \quad [\mathbf{R}_R]_{i,j} = \epsilon_R^{|i-j|}$, where \( 0 \leq \epsilon_T < 1,\; 0 \leq \epsilon_R < 1\) denote the transmit and receive correlation coefficient. Herein, we set $\epsilon_T = \epsilon_R = 0.8$ to characterize strongly correlated MIMO channels.

\begin{figure}[!t]
\centering
\subfloat{\includegraphics[width=0.495\columnwidth]{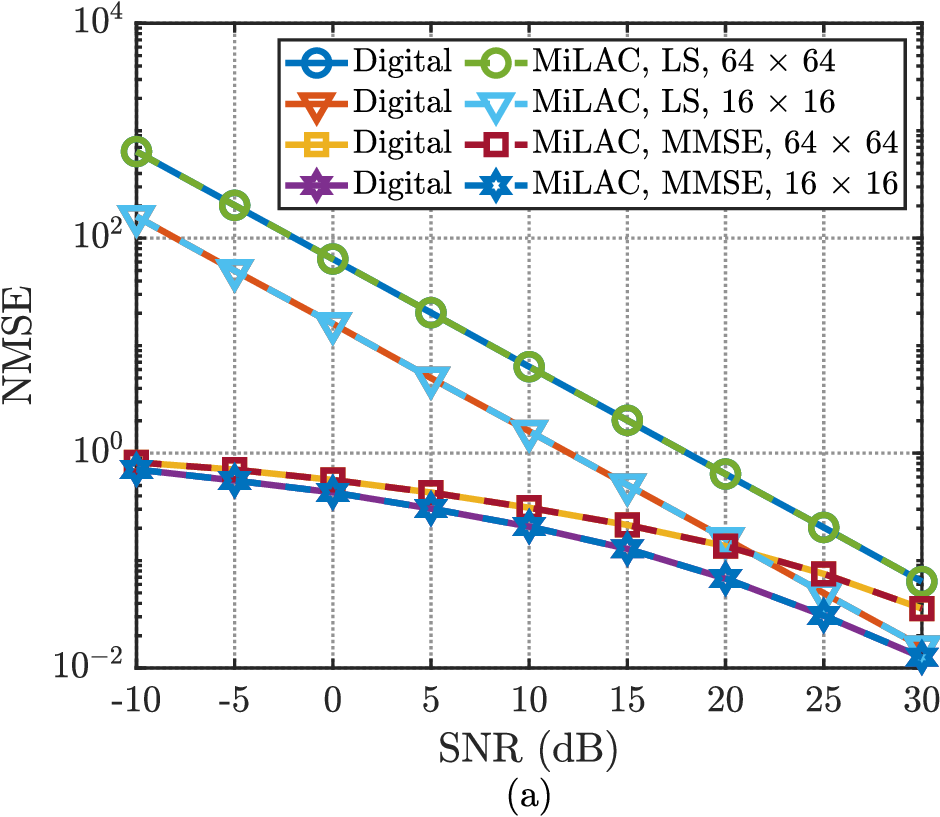}
\label{fig.2.a}}
\subfloat{\includegraphics[width=0.495\columnwidth]{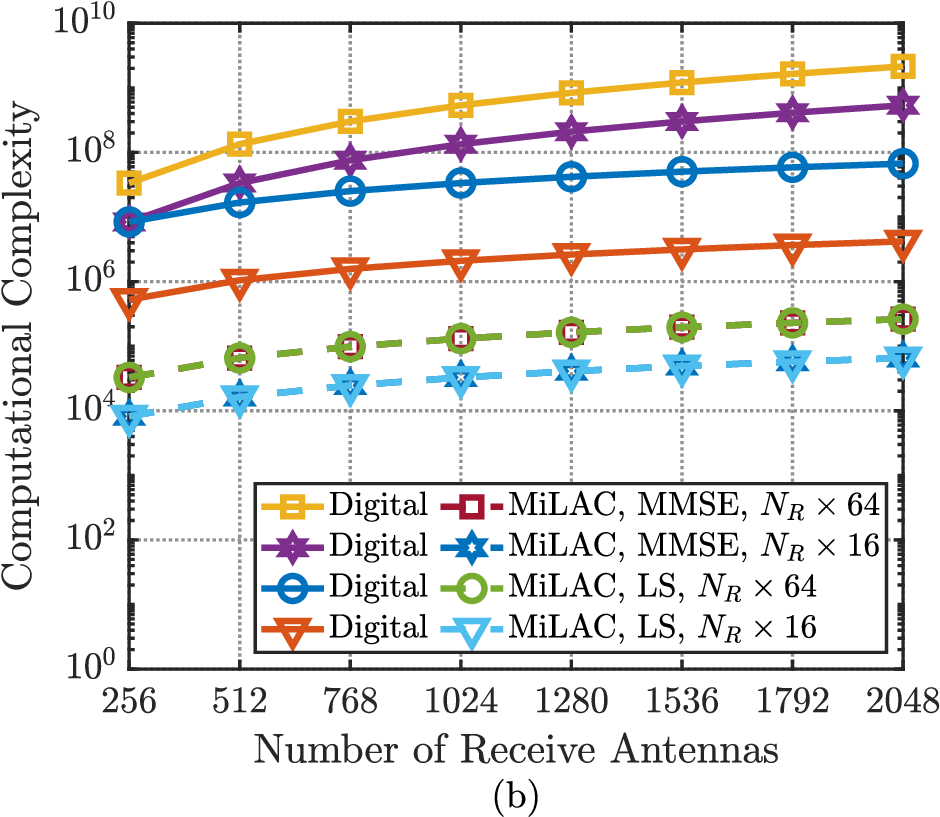}
\label{fig.2.b}}
\caption{Comparison of digital and MiLAC-aided channel estimation in an $N_R \times N_T$ MIMO system: (a) NMSE versus SNR; (b) computational complexity versus number of receive antennas.}
\label{fig.2}
\end{figure}

% Figure 2a
Fig.~\ref{fig.2.a} presents the $\text{NMSE} =\mathbb{E}\{\|\mathbf{H}-\hat{\mathbf{H}}\|_F^2\}/\mathbb{E}\{\|\mathbf{H}\|_F^2\}$ versus the \(\text{SNR} = \bar{P}_T/\sigma^2\) achieved by digital and MiLAC-aided LS and MMSE channel estimation, where $N_T = N_R \in \{16,64\}$. As expected, LS estimation suffers from higher NMSE, especially at low SNR, while MMSE estimation achieves improved performance by exploiting channel statistics. A slightly higher NMSE is observed for larger antenna numbers, as $\bar{P}_T$ is fixed. Notably, MiLAC-aided LS and MMSE achieve the same NMSE as their digital counterparts across varying antenna numbers, validating our theoretical insights in Sections \ref{Sec.LS.C} and \ref{Sec.MMSE.C}.

% Figure 2b
Fig.~\ref{fig.2.b} evaluates the computational complexity of digital and MiLAC-aided LS and MMSE channel estimation as a function of $N_R$, varying from $256$ to $2048$, with $N_T \in \{16,64\}$. We have the following observations. \textit{First}, digital MMSE requires more computation than digital LS, since its computational complexity $8\tau N_R^2$ grows quadratically with $N_R$, while digital LS scales as $8\tau N_RN_T$, linearly with $N_R$ but quadratically with $N_T$, given $\tau = N_T$. This explains why increasing $N_T$ results in a sharper computational complexity rise for digital LS. 
\textit{Second}, MiLAC-aided LS and MMSE achieve much lower computational complexity than their digital counterparts, with up to an $8192\times$ reduction, corresponding to a saving of up to $2.15\times10^9$ real operations, when MMSE estimation is applied in a $2048\times 64$ MIMO system. 

\section{Conclusion}
In this letter, we propose efficient LS and MMSE channel estimation schemes for MIMO systems 
aided by lossless and reciprocal MiLACs. These schemes perform LS and MMSE estimation, which otherwise require intensive digital computation, largely in the analog domain, achieving the same performance as their digital counterparts with significantly lower computational complexity. These results highlight MiLAC as a potential solution for efficient channel estimation in gigantic MIMO systems.
Future research directions include assessing the estimation performance of the proposed schemes under practical nonidealities, such as component tolerances and drift, insertion loss, impedance mismatch, and mutual coupling, as well as evaluating their computational complexity while accounting for practical implementation aspects, such as tuning and calibration. It will also be interesting to investigate channel estimation for MiLAC-aided wideband systems.

\appendix
\label{Appendix}
Following \cite[Appendix]{S3:4}, we define computational complexity as the number of real arithmetic operations performed in the digital domain per coherence block, considering no cost for matrix transpose/Hermitian, and two and six real operations for each complex addition/subtraction and multiplication, respectively. Under this convention, real-valued entry-wise scaling of a complex vector $\mathbf{a} \in \mathbb{C}^{N\times 1}$ requires approximately $2N$ real operations, a matrix-vector product $\mathbf{B}\mathbf{a}$ with complex $\mathbf{B} \in \mathbb{C}^{M\times N}$ requires approximately $8MN$ real operations, and a matrix-matrix product $\mathbf{B}\mathbf{A}$ with complex $\mathbf{A} \in \mathbb{C}^{N\times L}$ requires approximately $8LMN$ real operations.

\section*{Acknowledgment}
The authors thank the anonymous reviewers and the editor for their careful review and constructive comments.

\bibliographystyle{IEEEtran}
\bibliography{refs}

\vfill

\end{document}